 \documentclass[a4paper]{article}

\usepackage{fancyhdr}

\usepackage{graphicx}
\usepackage{stmaryrd}
\usepackage{amsmath,amssymb,amsthm}
\usepackage{empheq,soul}
\usepackage{color}
\usepackage{upgreek}
\usepackage{bbold}
\usepackage{hyperref}
\usepackage{etoolbox}
\usepackage{xspace}
\usepackage{tabularx}
\usepackage{booktabs}
\usepackage{caption}
\usepackage{tikz}
\usepackage[Q=yes]{examplep}

\usetikzlibrary{plotmarks}

\usepackage[final]{listings}
\lstdefinestyle{INLINE}{
  basicstyle=\tt,
}
\lstdefinestyle{DISPLAY}{
}
\lstdefinestyle{FLOAT}{
  float,
  captionpos=b,
  basicstyle=\small\tt,
  numbers=left,
  numberstyle=\footnotesize,
  numbersep=0.5em,
}
\lstdefinestyle{SFLOAT}{
  float,
  captionpos=b,
  basicstyle=\footnotesize\tt,
  numbers=left,
  numberstyle=\footnotesize,
  numbersep=0.5em,
}

\lstnewenvironment{JF}[1][]{\lstset{style=DISPLAY,style=FLOAT,#1}}{}
\lstnewenvironment{JSF}[1][]{\lstset{style=DISPLAY,style=SFLOAT,#1}}{}
\lstnewenvironment{JD}[1][]{\lstset{style=DISPLAY,#1}}{}

\newcommand{\dotr}{\mbox{${\cdot}$}}

\definecolor{highlight}{gray}{0.85}

\numberwithin{equation}{section}

\begin{document}

\title{Actors vs Shared Memory: two models at work\\ on Big Data
   application frameworks}

\author{Silvia Crafa\\
        Universit\`a di Padova, Italy
       \and Luca Tronchin \\
        Universit\`a di Padova, Italy}
\date{}
\maketitle

\begin{abstract}
This work aims at analyzing how two different concurrency models,
namely the shared memory model and the actor model, can influence the
development of applications that manage huge masses of data,
distinctive of Big Data applications.  
The paper compares the two models by analyzing a
couple of concrete projects based on the MapReduce and Bulk
Synchronous Parallel algorithmic schemes. Both projects are
doubly implemented on two concrete platforms: Akka Cluster and Managed
X10.  
The result is both a conceptual comparison of models 
in the Big Data Analytics scenario, and an experimental analysis based
on concrete executions on a cluster platform.
\end{abstract}




\section{Introduction}
\label{sec:introduction}
In recent years the concept of Big Data has attracted the general
interest since valuable information can be extracted from the analysis
of large volumes of data. Even if there is not a reference size, Big
Data generally refers to those data sets whose size requires hardware
and software tools beyond the common machines capabilities. 
A crucial aspect of Big Data computing is concurrency, since data must
be processed in parallel by distributing the computation across
multiple machines.
To facilitate this process, specific application frameworks are
becoming widespread as they autonomously manage the parallelism and
distribution of the operations, requiring the 
user to only define the processing of small amounts of data.
In this work we focus on a couple of algorithmic schemes that lie
under the major Big Data application frameworks, that is 
\textit{MapReduce} and the
\textit{Bulk Synchronous Parallel} model.  
MapReduce is a model inspired by \textit{map} and \textit{reduce}
functions from the functional programming, through which many real
world tasks can be expressed (e.g., sorting and searching tasks
\cite{petasort,goodrich_11_mapreduce-theory,DBLP:KajdanowiczKI13}).  
The Bulk Synchronous Parallel (BSP) model has recently gained great
interest because of its ability to process graphs of enormous size,
such as social or location graphs,
by deeply exploiting the work distribution on several
machines (\cite{Krizanc:1996:BSP:882471.883319,Bisseling:BSP,Valiant:2011:BMM:1889388.1889509}). 
These two algorithmic schemes are abstract enough to be implemented on
any platform; there are indeed several implementations based on the
most popular languages. We just mention here the major frameworks:
Google's MapReduce~\cite{Dean04mapreduce:simplified}, 
Apache Hadoop and 
Apache Spark 
for MapReduce, and Google's Pregel~\cite{Pregel} and 
Apache Giraph 
for BSP.  

This work aims at studying how specific implementations of MapReduce
and BSP take advantage of the platform on which they are implemented.
In particular, we address the crucial core of the programming
paradigm, that is the concurrency model that shapes the platform
design. We then focus on the \emph{actor model} and the 
\emph{shared memory model}, in order to assess how they affect the
development of Big Data applications and their performance. 

To concretely instantiate the shared memory model, we consider the X10
platform, which is specifically designed for scale-out distributed
programming with a partitioned global address space.
As for the actor model, we consider the Scala language and the Akka
Cluster library. Both platforms allow the development of applications
that run on the Java Virtual Machine (JVM), which thus provides a 
common basis to compare the applications and to bring out
the differences that specifically depend on each model. 
%
Even if established Big Data application frameworks are available both
for X10 and Scala technologies (e.g., \cite{M3R-X10} and
\cite{Zaharia:2010:SCC:1863103.1863113}),  
we implement from scratch the MapReduce and BSP frameworks in a
lightweight but effective way, so to better value and compare the
distinctive features of the two underlying programming models. 

For the same reason it is also necessary to identify some tractable
problem whose significance can be appreciated 
($i$) at the abstract level, when comparing concurrency models,
($ii$) at the structural level, when comparing the programming styles,
and ($iii$) at the experimental level, when comparing performances. We
then apply the MapReduce scheme to the sorting of a one-dimensional
distributed array spanning over a set of nodes, and we instantiate BSP
on the exploration of a distributed sparse graph. Both problems are
implemented twice: as an X10 application and as an Akka application.
In the MapReduce projects we focus on remote parallelism, that is we
let each node just host sequential computation, while in the BSP
projects we study both local and remote parallelization
opportunities. 

Both the code style comparison and the experimental results
attest that X10 excels with \emph{inter-node parallelism}, distinctive
of MapReduce, while Akka shines on \emph{intra-node concurrency}
better scaling to the higher local concurrency degree required by BSP.

\paragraph{Structure of the paper}
In Section~\ref{sec:platforms} we overview the main characteristics
of the two platforms under examination. 
In Section~\ref{sec:algoMR-BSP} we illustrate the MapReduce and BSP
algorithmic schemes together with their instantiation on the problems
of sorting a distributed array and exploring a sparse graph. 
In Section~\ref{sec:implementations} we report on the implementation
of the two problems both in X10 and Akka, going into a comparison of
the resulting code styles and drawing some insight about the impact of
the different concurrency models. In Section~\ref{sec:experiments} we
provide the details of the experimental comparison and we put forward
our assessment of the impact of the two concurrency models in the Big
Data Analytics scenario. Finally, we draw our conclusions in
Section~\ref{sec:conclusions} pointing to future research directions.

\section{The programming platforms}
\label{sec:platforms}

X10\cite{Ebcioglu04x10:programming,Charles:2005:XOA:1094811.1094852} 
is an open-source object-oriented programming language designed 
for productive high performance programming of parallel and 
multi-core computer systems.
It is designed around a partitioned global address space (PGAS): 
the computation is spread over a set of \emph{places}, each of which
holds data and hosts asynchronous \emph{activities} that operate on
local data. Objects residing in one place may contain references
(\verb+GlobalRef+s) to objects residing in other places. However, X10
enforces a strong locality property: an object's mutable state
cannot be accessed through a remote reference to that
object. Therefore the language provides the {\tt at(p)S} construct 
that permits the current activity to change its place of execution to
{\tt p}, execute {\tt S} at place {\tt p} and return, leaving behind
activities that may have been spawned during the execution of {\tt S}.
The values of variables used in {\tt S} but defined outside {\tt S}
are serialized, transmitted to {\tt p}, and de-serialized to
reconstruct a binding environment in which {\tt S} is executed.

Activities represent lightweight threads in X10 and can be created and
started locally with the {\tt async S} statement. 
Dually, the {\tt finish S} statement executes {\tt S} and waits
for the termination of every activity spawned by {\tt S} either
locally or at a remote place.
Given the underlying shared-memory model, atomic execution is provided
by the \verb+atomic S+ construct, which guarantees
that a single local activity at a time can enter
its atomic block.
Finally, X10 supports global data-structures spanning multiple
places. For instance, as described in
Section~\ref{sec:implementations}, we will take advantage of the
\verb+DistArray+ object, which represents  
a global reference to a distributed array and provides useful methods
that simplify the centralized control of its distributed elements.

As far as Akka is concerned, in this work we draw on the combination
of the Scala language and Akka's actor implementation  included 
in its standard library. 
Actors are objects that encapsulate state and behavior; they
communicate exclusively by asynchronously exchanging messages which
are placed into the recipient's mailbox.
The environment takes care of the actor mailbox and guarantees
that a single messages is processed at a time, hence as long
as only the actor has access to its local variables, the absence
of race conditions is enforced by the model.
%
As for distribution, we rely on the Akka Remote and Akka Cluster
libraries, which support referential transparency in remote
actors communications, that is the developer can use the actor
references (\verb+ActorRef+) to send messages without having to worry
about whether the recipient resides on the same node of the sender or
not.  
There are two main types of actor references: local actor references
and remote actor references. The remote references contain the 
information necessary to identify the actors on different nodes.
The actor references may be exchanged between actors through messages 
or can be calculated since an actor can be identified in the system 
by combining the \emph{node address} with the
\emph{actor path} in the node.
Finally, we mention that in order to monitor and manage the cluster
status, Akka Cluster relies 
on the gossip protocol to support the elastic and
resilient nature of the system. This component exchanges a
bunch of implicit messages that are not directly controlled by the
user, which however only cause a small overhead in actor
communications.

\section{Two BigData algorithmic schemes}
\label{sec:algoMR-BSP}

\subsection{MapReduce}

\begin{figure}[ht]
	\centering
	\fbox{
		\includegraphics[scale=0.49]{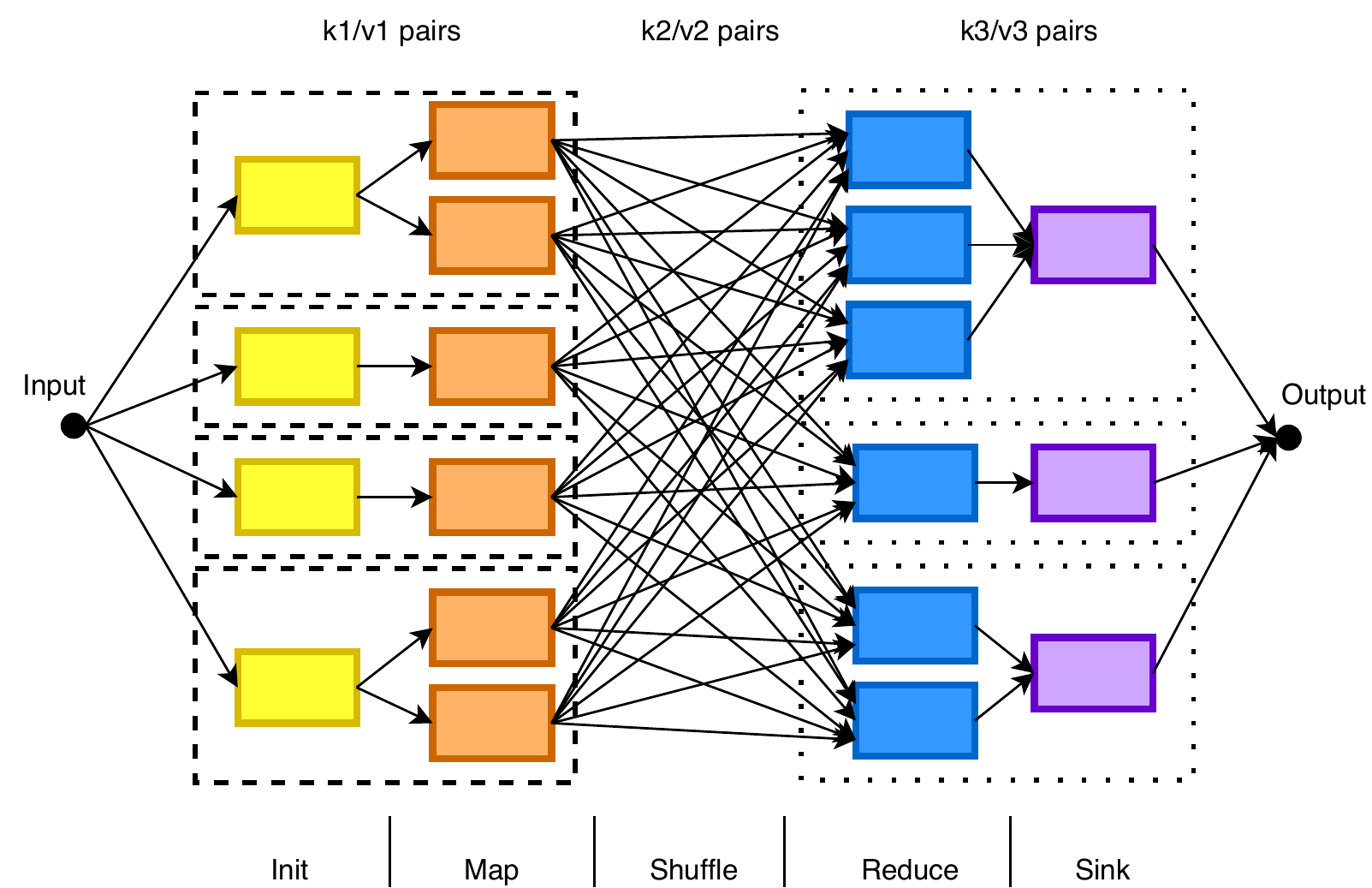}
		}
	\caption{Diagram of MapReduce.}
	\label{fig:mapreduceDiagram}
\end{figure}

In the MapReduce algorithmic scheme, a distributed computation is
specified as a (possibly repeated) sequence of \emph{Map},
\emph{Shuffle} and \emph{Reduce} steps, 
parametrized around a pair of functions supplied by the user.
The first function, \verb+mapper+, processes a (key,value)
pair to generate an intermediate (key,value) pair, and the
second function, \verb+reducer+, merges all the values associated
with the same intermediate key to output a final (key,value) pair. 
More precisely, the algorithm corresponds to the following
sequence of steps, depicted in Figure~\ref{fig:mapreduceDiagram}:
\begin{itemize}
  \item \textit{Initialization} step: data are loaded in the initial
    (key,value) pairs, in order to process them in the Map step.  
  \item \textit{Map} step: each node applies the \verb+mapper+
    function to the local (key,value) pairs obtaining intermediate
    pairs. 
  \item \textit{Shuffle} step: each node redistributes intermediate
    pairs produced by the \verb+mapper+ function, such
    that all values belonging to one key are located on the same 
    node. 
  \item \textit{Reduce} step: each node now processes each group
    of pairs, per key, in parallel, using the \verb+reducer+ function.
  \item \textit{Sink} step: it processes the Reduce output to apply
    some final operation like storing the results. 
\end{itemize}

As for the distribution, 
each node performs the \textit{Init} step independently, in parallel
to the others. Subsequently each node performs the \textit{Map} step
on data produced by the local \textit{Init} step (dashed line
boxes). 
In the same way, data processed by a \textit{Reduce} step are then
processed by a \textit{Sink} step on the same node (dotted line
boxes). 
As for the \textit{Shuffle} step, the user defines a
\verb+partition+ function to determine the target node depending on
the output key of the \textit{Map} step.

Figure~\ref{fig:mapreduceDiagram} shows that in each node
the \textit{Map} and \textit{Reduce} steps can be locally parallelized
by applying the \verb+mapper+ and \verb+reducer+ functions
to different portions of local data. 
Indeed, MapReduce can take advantage of \emph{locality} of data, by
processing them on or near the storage assets in order to reduce the
distance over which they must be transmitted. 
Therefore, at every step,
the nodes only work on their own data portion,
eliminating data races,
and local computation proceeds independently of the remote one.
Hence the change in the number of nodes does not involve the
computation, but only affects performance. 

While this process can often appear
inefficient compared to algorithms that are more sequential, or
designed to run on a single machine, MapReduce can be applied to
significantly larger data-sets than ``commodity'' servers can handle (a
large server farm can use MapReduce to sort a petabyte of data in only
a few hours\cite{petasort}). 
However,
 when designing a MapReduce algorithm, the author needs to choose a
good trade-off\cite{Ullman:2012:DGM:2331042.2331053} between the
computation costs and the communication costs of the shuffle
step. In particular, the partition function and the amount of data
transferred after the Map step can have a large impact on the
performance; indeed, in the Big Data scenario communication cost often
dominates the computation cost. 

\subsection{Sorting a distributed array with MapReduce}

\begin{figure}[ht]
 	\centering
 	\fbox{
 		\includegraphics[scale=0.46]{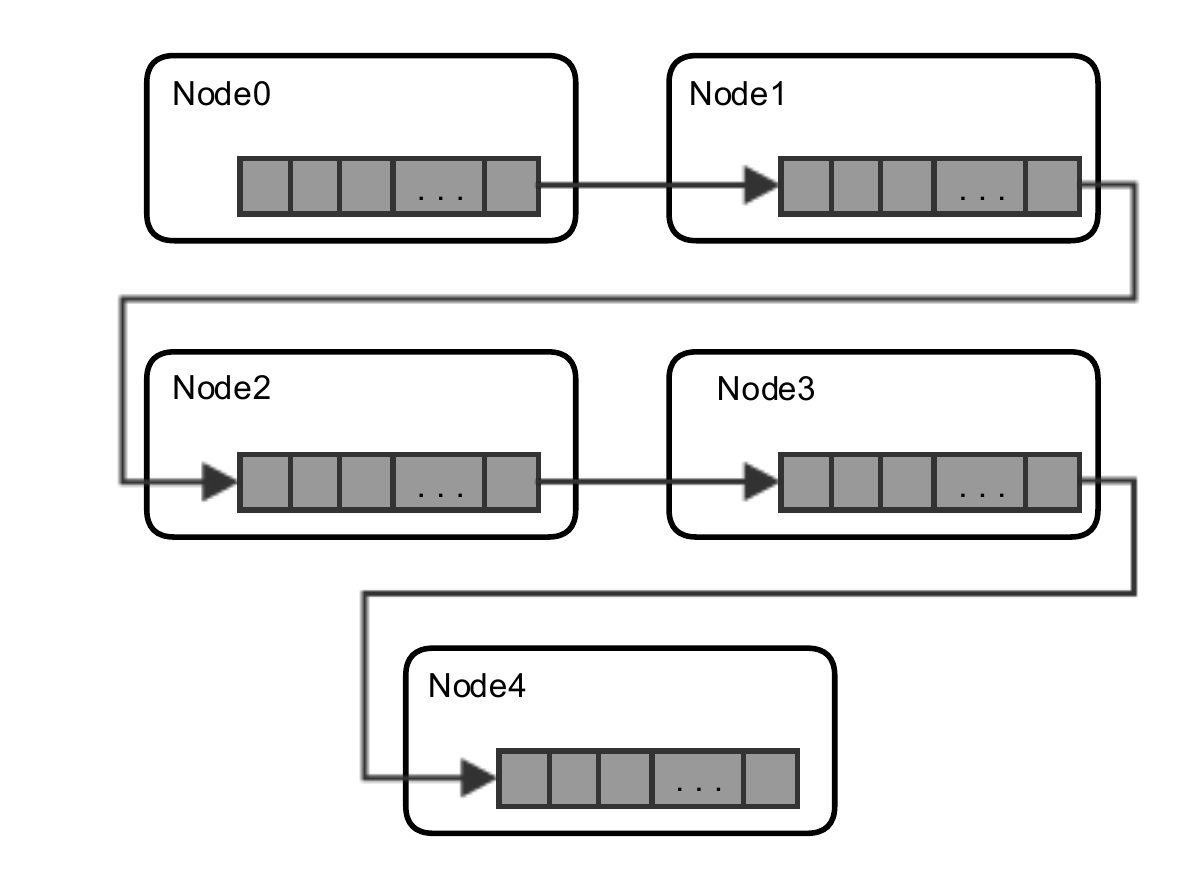}
 		}
 	\caption{An array spanning multiple nodes.}
 	\label{fig:distributedArray}
 \end{figure}

We now apply the MapReduce algorithmic scheme to the sorting of
a one-dimensional distributed array spanning over a set of nodes. 
We assume that the cluster has a fixed number of $N_{\mathit{nodes}}$ nodes, that are
ordered from $0$ to $N_{\mathit{nodes}}-1$. 
We let the value of the input array to be randomly generated 
and we let the array to be equally distributed over the nodes so that
the indexes of the array slice 
allocated on any node $p$ 
are 
greater than the indexes of the slice allocated on the node $p-1$, as
depicted in Figure~\ref{fig:distributedArray}.
%
We also require the output array to be sorted, to be equally
distributed over the nodes, and the elements at node $p$ must be
greater than those at node $p-1$, for any $p\in 1,...,N_{\mathit{nodes}}-1$. 

Before the MapReduce execution starts, there is a little
pre-elaboration where the global maximum and global minimum of the
whole array are retrieved, so to compute the $\mathit{Range}$ of the
values in the array.

We now define the behavior of every step. We denote by
\verb+k1+ and \verb+v1+ respectively the key and the value in input
for the \textit{Map} step; \verb+(k2,v2)+ is the output pair of
\textit{Map} and the input of the \textit{Reduce} step;
\verb+(k3,v3)+ is the output of the \textit{Reduce} step. 

\paragraph{Initialization step}
A \verb+Source+ function is performed once on each node to define the
(key,value) pairs to be passed to the \emph{Map} step. 
More precisely, for every element $i$ of the local array slice, it defines the key \verb+k1+$_i$ as the 
global position 
in the array and the value \verb+v1+$_i$ as the essential data needed
for comparing $i$ with other elements. 
In this case using ``essential data'' instead of the whole
element is convenient because \verb+v1+$_i$ might be too heavy to
be transferred from a node to another during the intermediate steps. 
Storing the original position in the key is essential to finally
transfer the whole element once its destination has been found.  

\paragraph{Map step}
This step applies the \verb+mapper+ function to each (local) element
$i$. In particular, the function computes a destination node for the
element $i$, storing it in the output key.
Let {\it min} be the global minimum; we split the range of values
into $N_{\mathit{nodes}}$ intervals, each of size
$I=\mathit{Range}/N_\mathit{nodes}$, 
and we intend to use each node $p$ to host values in the interval 
$$
I_p = [\mathit{min}+pI,\dots, \mathit{min}+(p+1)I-1]
$$
Then the \verb+mapper+ function takes in input the pair 
\verb+(k1+$_i$\verb+,v1+$_i$\verb+)+ and produces the output pair
\verb+(k2+$_i$\verb+,v2+$_i$\verb+)+ where
\verb+k2+$_i=q$ for some $q$ such that 
\verb+v1+$_i \in I_q$, and
\verb+v2+$_i=$\verb+(k1+$_i$\verb+,v1+$_i$\verb+)+.  

\paragraph{Shuffle step}
Given the mapper function above, the partition function used in the
shuffle step is the identity function. This means that each pair
\verb+(k2+\verb+,v2)+ is sent at node \verb+k2+.
Therefore, since the nodes are
ordered, at the end of this step we have that for any pair
of nodes $p$ and $q$, with $p < q$, all the elements located in $q$
are greater than all the elements located in $p$. On the other hand
the array slices hosted at different nodes are not sorted
and in general have different sizes. 

\paragraph{Reduce step}
As a consequence of the previous steps, at this stage all the pairs
hosted at each node share the same key \verb+k2+.
Hence the \verb+reducer+ function is performed once for every
different node to sort the local elements with the best algorithm.
At the end the local slice of the array is sorted and no 
elements will be inserted anymore. More precisely, the output of
\verb+reducer+ 
at node \verb+j+ is the pair \verb+(k3+$_j$\verb+,v3+$_j$\verb+)+,
where \verb+v3+$_j$ is the sorted slice and \verb+k3+$j$ its first
element.  

\paragraph{Sink step}
As observed before, the array slices sorted at different nodes have
different sizes. The final \verb+Sink+ step, performed once on each
node, is then necessary to equally redistribute the sorted array over
the nodes. All the sorted sub-arrays will be sent to a specific node,
which will take care of the elements redistribution.
Notice that the \verb+Sink+ step is logically simple, but involves a
huge data transfer between nodes and it is unavoidable.

\subsection{Bulk Synchronous Parallel}
\label{subsec:BSP}

Compared to MapReduce, BSP is best suited for the analysis of complex,
very-large-scale dynamic graphs (billions of nodes, billions of
edges). 
BSP programs consist of a sequence of iterations, called
\textit{phases}, and a set of logical units that communicate by means
of message passing. 
In order to avoid confusion with cluster nodes and actors, 
we refer to BSP logical units as \textit{Agents}.

\begin{figure}[ht]
	\centering
	\fbox{
		\includegraphics[scale=0.6]{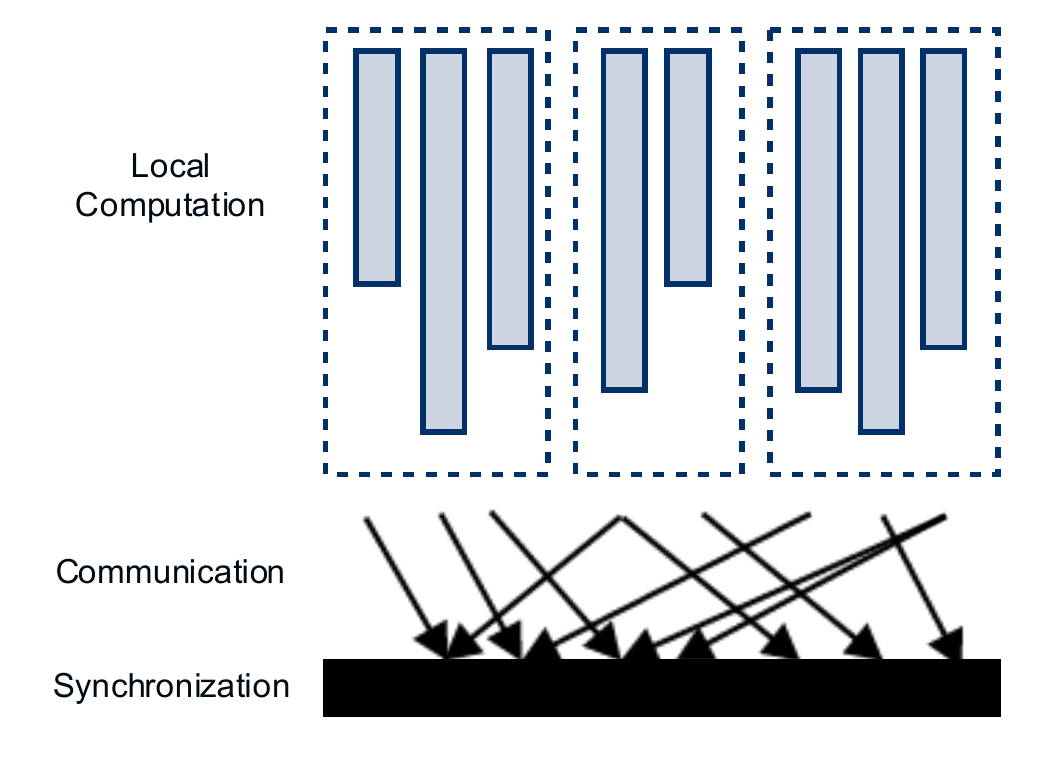}
		}
	\caption{Diagram of a BSP phase.}
	\label{fig:bspDiagram}
\end{figure}

Each phase $S$ is made of the following steps, 
depicted in Figure \ref{fig:bspDiagram} where dashed boxes
correspond to cluster nodes:
\begin{enumerate}
\item \emph{Local computation}: parallel agents compute a user-defined
  function, using their own 
  local memory. The function specifies the behavior of a single agent
  $V$ in the phase $S$: it can read messages sent to $V$ in
  phase $S-1$, send messages to other agents that will be received at
  phase $S+1$, and modify the state of $V$. 
\item \emph{Communication}: messages are exchanged between all
  agents.
\item \emph{Synchronization}: each node waits for the others to
  complete their message transfer at phase $S$, before starting phase
  $S+1$.
\end{enumerate}

In addition, each phase is characterized by a set of initially
\emph{active} agents. These agents execute local computation, then
deactivate themselves, and activate other agents by sending a
message to them. 
A \emph{job} that executes in a BSP machine is defined by a set of
agents, a subset of starting active agents, and a stop criterion. 
A BSP machine stops when there are no more active agents or when the
stop criterion is reached even if there are yet active agents.

Figure \ref{fig:bspDiagram} shows how each agent runs independently
and in parallel to the others, both local and remote, but each agent 
waits for the others at the 
barrier synchronization, before starting the next phase.

An important part of the analysis of a BSP algorithm rests on
quantifying the synchronization and communication that are needed. 
On the other hand, the BSP agents can be programmed by only focusing
on their local memory
without directly referring to remote entities.

\subsection{Sparse graph exploration with BSP}

We now describe how the BSP algorithmic scheme applies to the
exploration of a sparse graph. 
We consider a random directed graph generated with up to 3 edges for
every vertex. We assume that each node of the cluster hosts $N$
vertexes, and we let a vertex correspond to a BSP-agent. 
Every edge can point to a vertex in the same node or to a
vertex in a different node, but we assume that the whole graph
is \emph{simple}, that is there are no loops and no more than one
edge with the same source and destination.

Each vertex has a \verb+parent+ field and an inbox listing the
received messages. 
The local computation step, executed at every phase by every active
vertex, is so defined: 
the vertex checks the value of its \verb+parent+ field,
if it has a non-null value than it deactivates itself,
otherwise 
($i$) it sets its \verb+parent+ field with the content of the
first received message, 
($ii$) than it sends a message to every target vertex of any out-coming
edge, 
($iii$)	and finally deactivates itself.

Every vertex that receives a message will be active in the next phase.
At the beginning a start vertex is marked as active so to begin the
computation. In this problem there is no stop criterion, so the
execution stops when there are no more active agents.
The \verb+parent+ field and the barrier synchronization between each
phase guarantee that each vertex is visited only once, hence the
output is a tree. On the other hand, the resulting tree does not
contain the vertexes that have not been visited since they were not
reachable from the start vertex.

\section{Implementation and code style comparison}
\label{sec:implementations}
In this section we comparatively discuss our implementation of the
distributed array sorting and the sparse graph exploration problems,
both in X10 and Scala+Akka Cluster. 
First of all, we observe that even if established Big Data application
frameworks are available for both technologies (e.g., MR3 for
X10~\cite{M3R-X10} and Spark for Scala~\cite{Zaharia:2010:SCC:1863103.1863113}), we decided to
implement from scratch the MapReduce and BSP frameworks in a
lightweight but effective way, so to better value and compare the
distinctive features of the two underlying programming models. 
We assumed for simplicity that the cluster has a fixed number of
nodes, and if a node fails we let the whole computation fail.
Although Akka is elastic and resilient by nature, and also X10
supports a resilient mode~\cite{X10resilient}, we think that
resiliency is orthogonal w.r.t. our goals, and we postpone the
comparison of a resilient version of our applications to future work. 
We also observe that to get
best performance with X10 one should use Native X10, which compiles to
C++. However we are not interested in absolute performances, we aim
instead at drawing some insight about the impact  
of the concurrency model on the application performances. 
Hence we rely on Managed X10, which compiles to Java, so to better
compare with the Scala+Akka technology.

A first important remark is that, besides the different concurrency
models, 
the two programming platforms under examination significantly 
differ in their structure. As a consequence, even the programming 
styles that the developer must adopt, substantially differ. 
We illustrate below these differences, postponing to
Section~\ref{sec:experiments} the experimental comparison. 

\paragraph{Project structure}
Each one of the four programs is divided in two distinct parts:
($i$) the \emph{engine}, consisting of the modules that implement the
algorithmic scheme (MapReduce or BSP) parametrized
w.r.t. the job to be executed, 
and ($ii$) the \emph{job} instance (distributed sorting, graph
exploration), to be submitted to the engine.
This modular structure entails genericity and reusability, and allows
a more effective comparison between the two frameworks.
The projects based on Akka are formed by a larger
  number of \emph{modules} than those in X10. 
This because Akka entails a more fragmented 
structure, in which each actor embodies a specific functionality,
i.e. a certain role in the project. There are actors
representing worker nodes, distinguished actors acting as control
nodes, 
and a number of actors implementing the aggregator pattern to
coordinate interactions.
Instead, 
X10 \emph{views the cluster at an higher abstraction level} than
Akka, being able to describe in the same \emph{module} (i.e. block of
code) more computation steps which are usually separated in the most
popular programming languages. This is especially possible with
the language constructs distinctive of X10: \verb+at+,
\verb+async+, and \verb+finish+. 
In particular, the \verb+at+ construct considerably abstracts and simplifies the distribution of the computation on different nodes.
This allows to have a \textit{direct view of the distribution} 
to be obtained, which promotes code readability. 
Moreover, the joint use of these three constructs is very effective in
retrieving a value from all nodes or, in general, in the centralized
management of the global computation
(e.g., in the main classes of the MapReduce and BSP engines).

\paragraph{Main control flow and program launch}
The difference in the structure of the two projects is reflected on
the main control flow of the programs. In X10 the main control flow is
rendered as a control loop that asynchronously spawns code to be
executed in every place, while in Akka the control flow is spread on
several modules taking advantage of the aggregator pattern.

Even the procedure to launch the program execution reflects the
different nature of the two platforms. Starting an X10 execution
environment is not so different from starting any other mono-machine
program. Basically it requires that an SSH connection is available
between all the machines, then a start-up script, provided by the
framework, activates all the nodes. The script basically requires only
two parameters: the number of \emph{places} to be activated, and an
IP address list of the available machines. 
If the number of places is greater than the number of IP addresses,
then more than one place will be activated on some machines. 
This start mode allows to easily execute the program in batch mode,
which will be required by our tests, as we shall see in the next
section.

Akka Cluster has instead a \emph{dynamic and decentralized view of the
cluster}, which makes the startup more complex. 
There is more than one way to boot the whole system,
however, in order to start the tests in batch mode, a script has been
developed to have a centralized boot of the system. 
The script requires an ``\verb+ip:port+'' list in order to know where
to place the \emph{Akka nodes}, and an SSH connection available
between all the machines. The script connects to the machines and
boots up every single node.

\paragraph{Data structures}
Being distributed by nature, X10 has some built-in useful
distributed data structures. In this case, \verb+DistArray+ represents
a generic multidimensional array distributed over some places, and
provides a centralized control of the array. 
For instance, the class
has the method \verb+localIndices+ that returns a subset of array's
indexes accessible from the place where it has been called; remind
that all the other indexes cannot be accessed from that place.
To illustrate, the following code retrieves the maximum value of the
whole distributed array. In the second line, the \verb+finish+ keyword
waits for the termination of all the asynchronous activities into the
inner block. 
In the third line, an asynchronous activity is started in every
place. Then, in each place, the maximum is retrieved
among the local values and is returned to the asynchronous call. At
last, when the \verb+finish+ is satisfied, the \verb+MaxReducer+ is
evaluated and the maximum value is stored to the \verb+max+ value. 

{\small
\begin{verbatim}
val max =  
    finish(Reducible.MaxReducer[Long](Long.MIN_VALUE)){
       for(p in Place.places()) at(p) async {
          var localMax:Long = Long.MIN_VALUE;
          for (i in originArray.localIndices())
            if (originArray(i) > localMax) 
               localMax = originArray(i);
          offer localMax;
       } 
};
\end{verbatim}
}
This piece of code also clearly shows how the control flow can be
managed in X10 within few lines.

Akka design states that every node is independent
and there must not be a single-point of failure, therefore 
there is no distributed data structure spanning over the
nodes, but it is
required to independently manage individual slices on every node. 
In the case of the distributed array, we let \verb+DistArrayNodeActor+
hold, in every node, a slice of the distributed array and manage all
the operations on it. 
As to retrieve the global maximum/minimum of the whole array, a
\verb+MinMaxAggregator+ actor is instantiated, so that it asks to the 
worker nodes the maximum/minimum of their local slice. Then the actor
collects the responses and extracts the global maximum/minimum.  

As for the distributed graph processed by BSP, each node holds its
local vertexes in an array of size $N$, and
each vertex holds in turn an array of three remote references
to other vertexes, that represent the edges of the graph.

\paragraph{The cost of communication}
In Section~\ref{subsec:BSP} we have seen that the second step of the
BSP algorithm involves message transfer between all agents.
This communication step is accomplished in X10 by storing the list of
received messages directly into the (possibly remote) recipient. 
Since X10 is a language with a shared memory model, and the insertion
of messages in this list is allowed to all the active agents at the
current phase, it is necessary to control the insertion of messages in
the list with an \verb+atomic+ block. 

Although the \verb+atomic+ block is an heavy construct 
(\cite{X10-Brief}), in this case
its use is appropriate, given the enormous number of possible
senders. Other solutions, such as a memory slot for each possible
incoming message, are inapplicable. We observe that the Shuffle and
Sink steps of MapReduce do not suffer from the same problem since 
in their case the possibility of predicting the target destination of
the involved communications allows to reserve a reasonable amount of
memory slots to be filled without resorting to
\verb+atomic+.

In the Akka-based implementation, the BSP-inbox of an agent is
implemented with an actor, \verb+MessageReceiver+, to which is
delegated the function of receiving messages. 
The code of \verb+MessageReceiver+ is reported below: 
collects in a single block all messages received in a given phase 
(5th line) to be returned to the agent in the \emph{next} phase (8th
line): 
%
{\small
\begin{verbatim}
class MessageReceiver[S] extends Actor {
  val inc = Array(ArrayBuffer.empty[Any], ArrayBuffer.empty[Any])
  def receive = {
    case Message(phase, x) => inc(phase%2) += x
      sender() ! MsgReceived

    case GetInbox(phase) =>
      sender() ! inc(phase%2).toArray
      inc(phase%2).clear()
 }
}
\end{verbatim}
}

As a final remark about Akka's communication model, we observe that
Akka does not impose constraints on the size of the messages exchanged
between the actors, that is, the messages within the same actor system
can have any size. However, the connections between the cluster
nodes are based on Akka Remote that uses the UDP protocol, which in
turn does not support arbitrary large datagrams. Therefore, 
large messages such as those exchanged in the Sink step of the
MapReduce projects, have to be chunked, and then re-compacted, by the
programmer.

\paragraph{Local and remote parallelism}
In the implementations of MapReduce we decided to focus on remote
parallelism, hence each node just hosts sequential computation. 
More precisely, all the local executions of
the \verb+mapper+ functions (one for each local array element) run
sequentially, and the \verb+reducer+ function is performed once for
each node. Each step of the algorithm is then executed in parallel at
the level of cluster nodes.
 
On the other hand, in the BSP implementations we studied
both local and remote parallelization opportunities. In particular,
observe that the local 
computation involved in the first step of each BSP phase can be
parallelized (see Figure~\ref{fig:bspDiagram}). Indeed, in our Akka
implementation, each agent executes in parallel;
remind that we let $N$ agents run on each node.
On the other hand, when $N$ grows, X10 does not properly scale-up:
as we shall see in the next section, X10 executions with
a large number of agents per node ($N\!>\!4000$) give rise to a
runtime warning saying that 
they are running too many threads, and execution does not terminate. 
We will more precisely discuss this issue in
Section~\ref{sec:experiments}, we just observe here that the problem
comes form the usage of the {\tt atomic} construct, whose
implementation blocks both the lightweight
activity, i.e., the logical task, and the underlying worker thread
(\cite{X10-Brief}).
We thus developed two X10 versions of BSP:
X10P, where local agents are executed in parallel,
and X10S, where agents in the same node are executed sequentially.
Anyway, notice that the atomic block described above is necessary even
in the case where the local agents are sequentially run,  
since local agent runs in parallel to agents in other places.

\subsection{Comparison of models}

We conclude this section by summarizing the comments above 
into a more general picture. 
Our detailed description of the four applications illustrates the very
different programming styles involved by X10 and Scala. 
Such a difference reflects the fact the design principles of these
languages significantly diverge, despite being both very expressive.
%

As far as X10 is concerned, even if functional constructs are
available and the \verb+async+ construct greatly simplifies
asynchronous execution, the underlying shared-memory model and the
centralized control of the distributed computation bring about an
imperative programming style with critical accesses to shared memory.
In this sense X10 is predisposed to a more synchronous
programming, where concurrency involve (large) sets of activities
running \emph{in parallel} with few coordination
requests. Accordingly, we will see in the next section that X10 shines
on the MapReduce application, which essentially consists of sequences
of steps with \emph{inter-node parallelism}.

In Scala, sticking to the actor model is a choice, since shared
variables are still available. For instance, using the \emph{futures} 
requires the operations scheduled at the future completion not to
modify the internal state of the actor so to prevent interferences.
However, the functional flavour of the language
promotes the definition of well-defined behaviours to be delegated and
composed by means of actors and futures. Such a disaggregation suits a
dynamic and decentralized view of the cluster, which fosters the
distributed execution of asynchronous operations boosting 
the parallelism.  Accordingly, we will see that Akka's asynchronous
programming style shines on the scale-up, that is in presence of
\emph{intra-node concurrency}, 
where a large number of \emph{concurrent} activities can be
effectively coordinated without the constraints and the limitations of
synchronization constructs.

\section{Experimental comparison}
\label{sec:experiments}

In this section we provide the details of our experimental comparison,
and we discuss the execution performances of the projects under
consideration. 

To develop the four projects we used the following tools:
X10 language (v. 2.5.1) and X10 Development Tool IDE (v. 2.5.1) for
X10 projects, and
Scala language (v. 2.11.4), Akka-actor (v. 2.3.8), Akka-remote
(v. 2.3.8), Akka-cluster (v. 2.3.8), IntelliJ IDEA IDE (v. 14.0.2)
and Scala Build Tool (v. 0.3.7) for Akka projects.
To perform the tests we used the cluster of the Department of
Mathematics of the University of Padova. 
The table below lists the machines we used, each
one with the following minimum requirements: 
8 cores, 16GB RAM, Infiniband connection, and 
JVM 1.6 (OpenJDK).

{
\footnotesize
$$\begin{array}{ r l r }
\toprule
\# & \text{CPU} & \text{RAM} \\
\midrule
 1 & \text{2 x Quad-Core Intel(R) Xeon(R) X5460 @ 3.16GHz} & \text{32GB} \\
 2 & \text{2 x Quad-Core Intel(R) Xeon(R) E5520 @ 2.27GHz} & \text{32GB} \\
 3 & \text{2 x Quad-Core Intel(R) Xeon(R) E5520 @ 2.27GHz} & \text{32GB} \\
 4 & \text{2 x Quad-Core Intel(R) Xeon(R) E5520 @ 2.27GHz} & \text{32GB} \\
 5 & \text{2 x Six-Core Intel(R) Xeon(R) X5650 @ 2.67GHz} & \text{64GB} \\
 6 & \text{2 x Quad-Core Intel(R) Xeon(R) X5460 @ 3.16GHz} & \text{16GB}\\
 7 & \text{2 x Quad-Core Intel(R) Xeon(R) E5520 @ 2.27GHz} & \text{32GB}\\
 8 & \text{2 x Quad-Core AMD Opteron(tm) 2378 @ 2.40GHz} & \text{64GB} \\
 9 & \text{2 x Quad-Core Intel(R) Xeon(R) E5520 @ 2.27GHz} & \text{32GB} \\
10 & \text{2 x Eight-Core Intel(R) Xeon(R) E5-2680 @ 2.70GHz} & \text{256GB} \\
\bottomrule
\end{array}$$
}

The access to the cluster is regulated by the Torque Resource Manager
queuing system, which defines the framework for the aggregate use.
To use the cluster it is then necessary to submit to the queuing
system an executable that specifies the resource requirements in terms
of the number of required machines and the number of required cores
for each machine; then Torque 
starts the program execution by assigning an IP address to each
required core, depending on the current load of the system. 
When more than one core per machine has been requested, 
the same IP address will be assigned multiple times.

As anticipated in Section~\ref{sec:implementations}, 
the X10 platform provides a script which, given the list of IP
addresses, launches the executable on 
each machine and autonomously establishes the connections
between the nodes (i.e. X10 places). Akka instead requires a manual
script that computes the various
``\verb+ip:port+'' combinations and starts each node separately.
This works
well, given the relatively small size we achieved; for larger clusters
(200, 1000, or more nodes), we would have to load the nodes
interactively, and adapt the configuration of the gossip protocol.

In the case of X10, a place then corresponds to a JVM, 
and similarly, an Akka node corresponds to a
single JVM. 


\paragraph{Description of the experiments}
Given the random generation of the problems and the intrinsic
nondeterminism, 
the reported results reflect the average time of ten executions
for each instance. The small variability revealed by the
execution times of the same instance is an indication of the
reliability of the obtained results. 

To measure the performance of programs, we first considered varying
the problem size, then varying the number of nodes in the cluster. 
From now on, when we consider the number of nodes in the cluster, we
will use the notation $A$x$B$, where $A$ is the number of physical
machines used, and $B$ is the number of nodes on that machine. 
This to highlight that part of the communication between nodes can
take place within the same machine and does not pass through a real
network, which is clearly slower.
%

It is important to observe that, besides the cluster size, in the
performance analysis it
is crucial to consider the single machine load. For instance, a
cluster 8x4 is very different from a cluster 4x8: they have
the same number of nodes, but with a different distribution between
the machines, leading to different performances (as illustrated by the
plots in the following subsection).
More precisely, since most of the machines have eight cores, 
a machine hosting eight nodes, such as those in the 4x8
cluster, is somehow ``saturated''. Instead, when just four nodes run
on a eight-core machine, the corresponding four JVMs enjoy more
powerful resources.
This key observation highlights the fact that in a performance
comparison it is not enough to consider just the number of  
nodes in the cluster, but also the available resources 
guaranteed by the machines hosting the nodes.

\subsection{MapReduce}

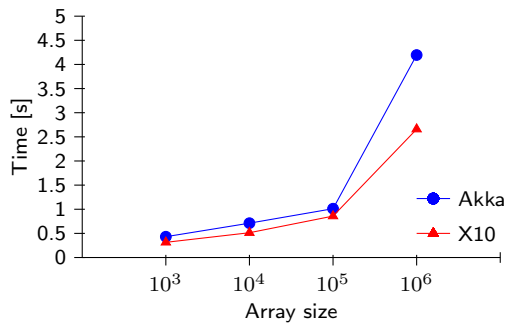
\begin{figure*}[t]
\footnotesize
\centering
\begin{tikzpicture}[y=0.64cm, x=1.1cm,font=\sffamily]
	\draw (0,0) -- coordinate (x axis mid) (5,0);
    \draw (0,0) -- coordinate (y axis mid) (0,5);

	\draw (1,1pt) -- (1,-3pt) node[anchor=north] {$10^3$};
 	\draw (2,1pt) -- (2,-3pt) node[anchor=north] {$10^4$};
 	\draw (3,1pt) -- (3,-3pt) node[anchor=north] {$10^5$};
 	\draw (4,1pt) -- (4,-3pt) node[anchor=north] {$10^6$};
	\draw (5,2pt) -- (5,-2pt);
    	\foreach \y in {0,0.5,...,5}
     		\draw (1pt,\y) -- (-3pt,\y) 
     			node[anchor=east] {\y};

	\node[below=0.5cm] at (x axis mid) {Array size};
	\node[rotate=90, above=0.55cm] at (y axis mid) {Time [s]};

	\draw[blue] plot[mark=*] 
		file {MR_arrayVar_Akka_10x3.data};
	\draw[red] plot[mark=triangle*]
		file {MR_arrayVar_X10_10x3.data};
    
	\begin{scope}[shift={(4,0.5)}] 
		\draw[red] (0,0) -- 
			plot[mark=triangle*] (0.2,0) -- (0.4,0)
			node[right,black]{X10};
	\end{scope}
	\begin{scope}[shift={(4,0.7)}]
		\draw[blue, yshift=\baselineskip] (0,0) -- 
			plot[mark=*] (0.2,0) -- (0.4,0) 
			node[right,black]{Akka};
	\end{scope}
\end{tikzpicture}
\caption{MapReduce - Variation of the array size on 10x3 nodes.}
\label{fig:MRarrayVar}
\end{figure*}

\paragraph{Variation of the array size}
Figure \ref{fig:MRarrayVar} shows the execution times of Akka
and X10 on 
the MapReduce implementation in a cluster where the
number of nodes is fixed (10x3), while the size of the array
varies from $10^3$ 
to $10^6$ integers.
What you notice is that Akka is a bit slower than X10 in all the
considered cases. Remind that in our implementation the local
computation is sequential, hence the program is locally synchronous.
We then think that the difference in the performances comes
from Akka's overhead due to asynchronous message exchange.


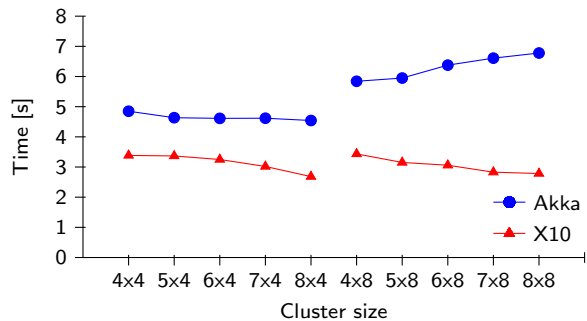
\begin{figure*}[t]
\footnotesize
\centering
\begin{tikzpicture}[y=0.4cm, x=0.6cm,font=\sffamily]
	\draw (0,0) -- coordinate (x axis mid) (11,0);
    \draw (0,0) -- coordinate (y axis mid) (0,8);

 	\draw (1,1pt) -- (1,-3pt) node[anchor=north] {4x4};
 	\draw (2,1pt) -- (2,-3pt) node[anchor=north] {5x4};
 	\draw (3,1pt) -- (3,-3pt) node[anchor=north] {6x4};
 	\draw (4,1pt) -- (4,-3pt) node[anchor=north] {7x4};
 	\draw (5,1pt) -- (5,-3pt) node[anchor=north] {8x4};


 	\draw (6,1pt) -- (6,-3pt) node[anchor=north] {4x8};
 	\draw (7,1pt) -- (7,-3pt) node[anchor=north] {5x8};
 	\draw (8,1pt) -- (8,-3pt) node[anchor=north] {6x8};
 	\draw (9,1pt) -- (9,-3pt) node[anchor=north] {7x8};
 	\draw (10,1pt) -- (10,-3pt) node[anchor=north] {8x8};
	\draw (11,2pt) -- (11,-2pt);
	\foreach \y in {0,1,...,8}
 		\draw (1pt,\y) -- (-3pt,\y) 
 			node[anchor=east] {\y};

	\node[below=0.5cm] at (x axis mid) {Cluster size};
	\node[rotate=90, above=0.55cm] at (y axis mid) {Time [s]};

	\draw[blue] plot[mark=*] 
		file {MR_clusterVar_Akka_1000000_4cores.data};
  	\draw[blue] plot[mark=*] 
    	file {MR_clusterVar_Akka_1000000_8cores.data};
	
  	\draw[red] plot[mark=triangle*]
			file {MR_clusterVar_X10_1000000_4cores.data};
  	\draw[red] plot[mark=triangle*]
    	file {MR_clusterVar_X10_1000000_8cores.data};
    
	\begin{scope}[shift={(9,0.8)}] 
		\draw[red] (0,0) -- 
			plot[mark=triangle*] (0.35,0) -- (0.7,0)
			node[right,black]{X10};
	\end{scope}
	\begin{scope}[shift={(9,1)}] 
		\draw[blue, yshift=\baselineskip] (0,0) -- 
			plot[mark=*] (0.35,0) -- (0.7,0) 
			node[right,black]{Akka};
	\end{scope}
\end{tikzpicture}
\caption{MapReduce - 
Variation of the cluster size with an array of $10^{6}$ elements.}
\label{fig:MR-Exp}
\end{figure*}

\paragraph{Variation of the cluster size}
The plot in Figure \ref{fig:MR-Exp} shows the comparison 
made by keeping constant the array size (set to
$10^6$ elements), while varying the number of nodes
in the cluster. Also in this case Akka is a bit slower, in a
uniform manner, compared to X10.
Notice in particular the central gap between the execution times in
a 8x4 and a 4x8 cluster: both programs have better
performances on ``non-saturated'' machines.
More generally, what emerges is that
both languages very well succeed in managing the change in the number 
of nodes, but the divergent directions of the segments 
might suggest that X10 scales-out better than Akka for larger
clusters. 
However, the considerations made above about the saturation of the 
machines indicate that no proper scale-out considerations can be
done without further tests.
For instance, the divergence of Akka might as well be explained
again by the overhead of the asynchronous management of messages,
aggravated by the machine saturation.

\subsection{Bulk Synchronous Parallel}

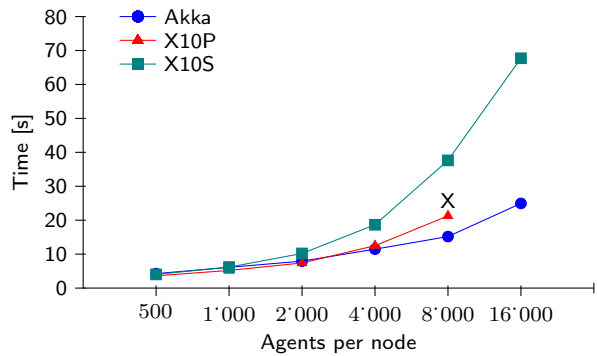
\begin{figure*}[th]
\footnotesize
\centering
\begin{tabular}{c}
\begin{tikzpicture}[y=.045cm, x=0.96cm,font=\sffamily]
	\draw (0,0) -- coordinate (x axis mid) (7,0);
    \draw (0,0) -- coordinate (y axis mid) (0,80);

	\draw (1,1pt) -- (1,-3pt) node[anchor=north] {$500^{ }$};
 	\draw (2,1pt) -- (2,-3pt) node[anchor=north] {$1^{\dotr}000$};
 	\draw (3,1pt) -- (3,-3pt) node[anchor=north] {$2^{\dotr}000$};
 	\draw (4,1pt) -- (4,-3pt) node[anchor=north] {$4^{\dotr}000$};
 	\draw (5,1pt) -- (5,-3pt) node[anchor=north] {$8^{\dotr}000$};
 	\draw (6,1pt) -- (6,-3pt) node[anchor=north] {$16^{\dotr}000$};
	\draw (7,2pt) -- (7,-2pt);
	
	\foreach \y in {0,10,...,80}
 		\draw (1pt,\y) -- (-3pt,\y) 
 			node[anchor=east] {\y};

	\node[below=0.5cm] at (x axis mid) {Agents per node};
	\node[rotate=90, above=0.55cm] at (y axis mid) {Time [s]};

	\draw[blue] plot[mark=*] 
		file {BSP_agentVar_Akka_8x4.data};
	\draw[red] plot[mark=triangle*]
		file {BSP_agentVar_X10P_8x4.data};
	\draw[teal] plot[mark=square*]
		file {BSP_agentVar_X10S_8x4.data};
    
	\draw(5,21.271) -- 
	 	plot[mark=triagle*, mark options={fill=white}] (5,21.271)
	 	node[above]{X};

	\begin{scope}[shift={(0.5,66)}] 
		\draw[teal] (0,0) -- 
			plot[mark=square*] (0.25,0) -- (0.5,0)
			node[right,black]{X10S};
	\end{scope}
	\begin{scope}[shift={(0.5,73)}] 
			\draw[red, yshift=\baselineskip] (0,0) -- 
			plot[mark=triangle*] (0.25,0) -- (0.5,0)
			node[right,black]{X10P};
	\end{scope}
	\begin{scope}[shift={(0.5,80)}] 
		\draw[blue, yshift=2\baselineskip] (0,0) -- 
			plot[mark=*] (0.25,0) -- (0.5,0) 
			node[right,black]{Akka};
	\end{scope}
\end{tikzpicture}
\\ \\
\begin{tabular}{c}
$\begin{array}{ l | *{6}{r} }
\toprule
\text{\#Agents} & 500 & 1^{\dotr}000 & 2^{\dotr}000 & 4^{\dotr}000 & 8^{\dotr}000 & 16^{\dotr}000\\
\midrule
\text{Akka} & 4.261 & 6.118 & 7.939 & 11.467 & 15.178 & 24.947 \\
\text{X10P} & 3.616 & 5.199 & 7.406 & 12.487 & ^* 21.271 & \text{N/A} \\
\text{X10S} & 4.029 & 6.175 & 10.201 & 18.660 & 37.628 & 67.718 \\
\bottomrule
\end{array}
$
\end{tabular}
\end{tabular}
\caption{BSP - Variation of agents per node on a cluster with 8x4 nodes.}
\label{fig:BSPagentVar}
\end{figure*}

\paragraph{Variation of the number of agents per node}
As we said in Section~\ref{sec:implementations}, we developed two X10
versions of BSP: X10P, where local agents are executed in parallel,
and X10S, where agents in the same node are executed sequentially. 
Although X10P can't handle all tested instances, it is reported to
actually compare X10 and Akka on the algorithm with the same degree
of parallelism at least on those instances in which X10P properly
terminates. 


Figure \ref{fig:BSPagentVar} shows a comparison of the
implementations of the BSP model, considering a cluster of constant
size ($8$x$4$), where the variation is made on the number of agents
per node. 
The plot and the table 
below with detailed time results, highlight how
Akka is slightly slower for small instances,
however its asynchronous model becomes more efficient in large
instances, because it avoids bottlenecks caused by the use of
\verb+atomic+. 
Also observe that the locally sequential version X10S is very slow, because
it does not exploit the parallel execution of the independent agents. 
The X10P case is the most efficient in smallest instances, where
concurrency on shared variables is lower, while it is worse than Akka
when the concurrency increases. 

%
The point in the plot labeled by $X$, that is the X10P instance with
$8000$ agents per node, corresponds to the average
between only two iterations (instead of ten), since the other eight
failed by suspending with a \emph{``too many threads''} warning. 
Instead, in the case of $16000$ agents per node, no execution
succeeded.

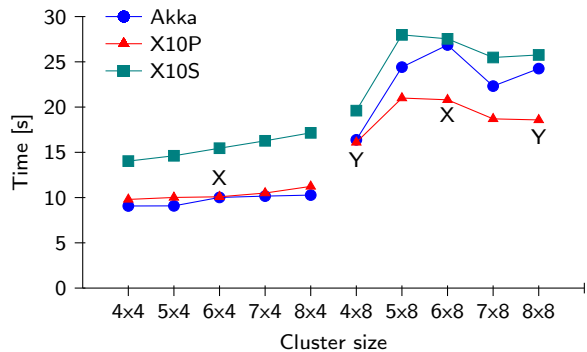
\begin{figure*}[th]
\footnotesize
\centering
\begin{tabular}{c}
\begin{tikzpicture}[y=.12cm, x=.6cm,font=\sffamily]
	\draw (0,0) -- coordinate (x axis mid) (11,0);
    \draw (0,0) -- coordinate (y axis mid) (0,30);

	\draw (1,1pt) -- (1,-3pt) node[anchor=north] {4x4};
 	\draw (2,1pt) -- (2,-3pt) node[anchor=north] {5x4};
 	\draw (3,1pt) -- (3,-3pt) node[anchor=north] {6x4};
 	\draw (4,1pt) -- (4,-3pt) node[anchor=north] {7x4};
 	\draw (5,1pt) -- (5,-3pt) node[anchor=north] {8x4};


 	\draw (6,1pt) -- (6,-3pt) node[anchor=north] {4x8};
 	\draw (7,1pt) -- (7,-3pt) node[anchor=north] {5x8};
 	\draw (8,1pt) -- (8,-3pt) node[anchor=north] {6x8};
 	\draw (9,1pt) -- (9,-3pt) node[anchor=north] {7x8};
 	\draw (10,1pt) -- (10,-3pt) node[anchor=north] {8x8};
	\draw (11,2pt) -- (11,-2pt);
	\foreach \y in {0,5,...,30}
 		\draw (1pt,\y) -- (-3pt,\y) 
 			node[anchor=east] {\y};

	\node[below=0.5cm] at (x axis mid) {Cluster size};
	\node[rotate=90, above=0.55cm] at (y axis mid) {Time [s]};

	\draw[blue] plot[mark=*] 
		file {BSP_clusterVar_Akka_4000a_4cores.data};
	\draw[blue] plot[mark=*] 
		file {BSP_clusterVar_Akka_4000a_8cores.data};
		
	\draw[red] plot[mark=triangle*]
		file {BSP_clusterVar_X10P_4000a_4cores.data};
	\draw[red] plot[mark=triangle*]
		file {BSP_clusterVar_X10P_4000a_8cores.data};
		
	\draw[teal] plot[mark=square*]
		file {BSP_clusterVar_X10S_4000a_4cores.data};
	\draw[teal] plot[mark=square*]
		file {BSP_clusterVar_X10S_4000a_8cores.data};

	\draw(3,10.5) -- 
	 	plot[mark=triagle*, mark options={fill=white}] (3,10.5)
	 	node[above]{X};
	\draw(6,12.5) -- 
	 	plot[mark=triagle*, mark options={fill=white}] (6,12.5)
	 	node[above]{Y};
	\draw(8,17.5) -- 
		plot[mark=triagle*, mark options={fill=white}] (8,17.5)
		node[above]{X};
	\draw(10,15) -- 
		plot[mark=triagle*, mark options={fill=white}] (10,15)
		node[above]{Y};

	\begin{scope}[shift={(0.5,24)}] 
		\draw[teal] (0,0) -- 
			plot[mark=square*] (0.4,0) -- (0.8,0)
			node[right,black]{X10S};
	\end{scope}
	\begin{scope}[shift={(0.5,27)}] 
		\draw[red, yshift=\baselineskip] (0,0) -- 
			plot[mark=triangle*] (0.4,0) -- (0.8,0)
			node[right,black]{X10P};
	\end{scope}
	\begin{scope}[shift={(0.5,30)}] 
		\draw[blue, yshift=2\baselineskip] (0,0) -- 
			plot[mark=*] (0.4,0) -- (0.8,0)
			node[right,black]{Akka};
	\end{scope}
\end{tikzpicture}
\\ \\
\begin{tabular}{c}
$
\begin{array}{ l | *{6}{r} }
\toprule
\text{\#Nodes} & \text{4x4} & \text{5x4} & \text{6x4} & \text{7x4} & \text{8x4}\\
\midrule
\text{Akka} & 14.248 & 14.530 & 16.656 & 16.319 & 16.125 \\
\text{X10S} & 31.831 & 35.507 & 36.801 & 39.560 & 40.693 \\
\bottomrule
\end{array}$
\\
$\begin{array}{ l | *{6}{r} }
\toprule
\text{\#Nodes} & \text{4x8} & \text{5x8} & \text{6x8} & \text{7x8} & \text{8x8}\\
\midrule
\text{Akka} & 25.787 & 25.490 & 25.665 & 28.116 & 28.043 \\
\text{X10S} & 47.097 & 51.036 & 51.935 & 53.704 & 58.379 \\
\bottomrule
\end{array}
$
\end{tabular}
\end{tabular}
\caption{BSP -
Variation of the cluster size with $4000$ agents per node.}
\label{fig:BSP-Exp}
\end{figure*}

\paragraph{Variation of the cluster size}
The  plot in Figure \ref{fig:BSP-Exp} considers the variation 
of the cluster size in BSP assuming a graph with $4000$ agents 
per node. Also in this experiment we distinguish the case of 
saturated and non-saturated machines.
In the tests with non-saturated machines the
executions of X10P and Akka last substantially the same time. 
X10S is instead proportionally slower than the others, again because 
of the lack of local parallelism.

On the other hand, on saturated machines, that is in the rightmost
part 
of the plot, there is a noticeable deterioration of execution times.
Observe that in this case Akka's performances are always in between
those of X10P and X10S, however the lack of linearity in the
results indicates that it is important to consider, also in this case,
the saturation of the machines.
As for X10P executions, one iteration failed in the cases 
6x4 and 6x8 (labeled with $X$), while four iterations 
failed in cases 4x8 and 8x8 (labeled with $Y$); this 
is a sign that the termination is not guaranteed.

In order to fully test Akka's scalability, we also tested 
the BSP program by assuming a graph with $10^4$ agents per node and
varying the cluster size. As illustrated by the execution times
reported in the table in Figure~\ref{fig:BSP-Exp}, 
Akka's implementation scales with the
size of the cluster and the corresponding increase of the number of
total agents. This then means that it scales well with the
proliferation of messages exchanged in the system.  
The table also shows that the execution time increase of X10S
is sharper, on the other hand, no X10P execution
terminated. 

\subsection{Performance analysis}

In the light of what emerged from the performance comparison,
we have that X10 excels with \emph{inter-node parallelism},
while Akka shines on \emph{intra-node concurrency}. It is important to
remind here that fair performance tuning X10 applications requires
different strategies (see~\cite{X10PerformanceTuning}), such that
avoiding the \verb+atomic+ construct and relying on Native X10 which
compiles to C++. 
Nevertheless, the main explanation of these results essentially
comes from the underlying programming models, as observed in
Section~\ref{sec:implementations}, which are indeed the main target of
our investigation.  

X10's centralized control appears to be best suited to inter-node
parallelism, hence to algorithms, like MapReduce, characterized by
sequences of independent steps with rare and predictable
synchronizations that minimize the need to protect the access to
shared memory. 
Instead, Akka makes the most of his asynchronous model in presence of
a high degree of concurrent activities, that can be efficiently
coordinated with no need of blocking mechanisms.

Let's consider this point more in depth: observe that both Akka and
X10 distinguish logical tasks (respectively the actors in Akka and the
activities in X10) from the worker threads, so to allow the execution
of many tasks with a few threads. 
However, in the two programming models such a distinction has a very
different impact. 
Indeed, consider how the X10 runtime realizes the decoupling between
tasks and worker threads: if a thread is about to block because of a
pending synchronization such as an \verb+atomic+ block, 
before suspending it starts a new worker thread with the aim of
preserving the parallelism. However the system has a limit (set by
default to $1000$ threads), after which the user is warned with the
already seen \emph{``too many threads''} message. 
Instead in the asynchronous model of actors the
tasks never block, hence worker threads can be mainly reused making
the best of the initial pool of threads. 
Moreover, \cite{OderskyHaller09}
illustrates how the functional model provides 
opportunities  to further
improve the decoupling between logical tasks and worker threads
by means of the continuation-passing style.
Anyway, despite the non-blocking implementation strategies, 
synchronous primitives like \verb+atomic+, 
are also \emph{logical blockers} that establish a conceptual
limit to the scale-up:
if a shared variable undergoes many synchronized accesses,
any task that needs to access must wait his turn, so the global
computation hardly advance. 
This is indeed what happens in the inbox of the BSP-agents, 
where a huge number of senders is ready
to deliver a message but few tend to do it.

\section{Conclusions}
\label{sec:conclusions}
In this paper we compared concrete implementations of the MapReduce
and BSP algorithmic schemes using the X10 and Akka Cluster platforms.
Rather than addressing performance issues, we aimed at testing the
actor model and the shared memory model at work on a Big Data
Analytics scenario. The experimental tests assess the expected
conceptual scale-up limit entailed by the blocking constructs required
to safely access shared memory.
Moreover, both the code style comparison and the
experimental results, attest that the centralized and imperative
flavour of X10 stands out in the MapReduce implementation, while
Akka's actors foster the distributed execution of asynchronous tasks
better scaling to the higher concurrency degree required by BSP.

As for the scale-out, 
we showed that both platforms can properly 
handle executions that involve up to 64 nodes. However, a more
faithful analysis requires additional tests with a larger cluster and
carefully tuning the resources available in each machine; this kind of
analysis is the subject of future work. 
We also plan to extend our investigation to assess the impact of the
programming models on fault tolerant executions. Resiliency is indeed
a critical issue of Big Data applications and both X10 and Akka
provide support for resilient executions.
In this case however it is necessary to suitably structure the
problems in order to evaluate the fault tolerance potential of the two
platforms.

\bibliographystyle{abbrv}

\begin{thebibliography}{10}

\bibitem{Bisseling:BSP}
R.~H. Bisseling and W.~F. McColl.
\newblock Scientific computing on bulk synchronous parallel architectures.
\newblock In {\em Proc. of IFIP 13th World Computer Congress}:(1), pages 509--514, 1994.

\bibitem{Charles:2005:XOA:1094811.1094852}
P.~Charles, C.~Grothoff, V.~Saraswat, C.~Donawa, A.~Kielstra, K.~Ebcioglu,
  C.~von Praun, and V.~Sarkar.
\newblock X10: An object-oriented approach to non-uniform cluster computing.
\newblock In {\em Proc. of OOPSLA'05}, pages 519--538, 2005. ACM.

\bibitem{X10resilient}
D.~Cunningham, D.~Grove, B.Herta, A.Iyengar, K.~Kawachiya, H.~Murata,
V.~Saraswat, M.~Takeuchi, and O.Tardieu.
\newblock  Resilient X10: efficient failure-aware programming.
\newblock  In {\em Proc. of PPoPP '14}, pages 67-80, 2014. ACM. 
 
\bibitem{petasort}
G.~Czajkowski, M.~Dvorsk\'y, J.~Zhao, and M.~Conley.
\newblock Sorting petabytes with {MapReduce} - the next episode.
\newblock
  \url{http://googleresearch.blogspot.com/2011/09/sorting-petabytes-with-mapreduce-next.html},
  2011.

\bibitem{Dean04mapreduce:simplified}
J.~Dean and S.~Ghemawat.
\newblock Mapreduce: simplified data processing on large clusters.
\newblock In {\em Proc. of OSDI'04}:(6), pages 10-10. USENIX Association, 2004.

\bibitem{Ebcioglu04x10:programming}
K.~Ebcioglu, V.~Saraswat, and V.~Sarkar.
\newblock X10: Programming for hierarchical parallelism and non-uniform data
  access.
\newblock In {\em Proceedings of the International Workshop on Language
  Runtimes, OOPSLA}, 2004.


\bibitem{goodrich_11_mapreduce-theory}
M.~T. Goodrich, N.~Sitchinava, and Q.~Zhang.
\newblock Sorting, searching, and simulation in the mapreduce framework.
\newblock In {\em Algorithms and Computation}: (7074), pages 374--383.
  Springer Berlin Heidelberg, 2011.

\bibitem{OderskyHaller09}
P. Haller and M. Odersky
\newblock Scala Actors: Unifying thread-based and event-based programming. 
\newblock \emph{Theoretical Computer Science}: 410(2-3), pages 202-220, 2009.

\bibitem{DBLP:KajdanowiczKI13}
T.~Kajdanowicz, P.~Kazienko, and W.~Indyk.
\newblock Parallel processing of large graphs.
\newblock {\em CoRR}, abs/1306.0326, 2013.

\bibitem{Krizanc:1996:BSP:882471.883319}
D.~Krizanc and A.~Saarimaki.
\newblock Bulk synchronous parallel: Practical experience with a model for parallel computing.
\newblock In {\em Proc. of PACT '96}, pages 208-217, 1996. IEEE Computer Society.

\bibitem{Pregel}
G.~Malewicz, M.~H. Austern, A.~J. Bik, J.~C. Dehnert, I.~Horn, N.~Leiser, and
  G.~Czajkowski.
\newblock Pregel: A system for large-scale graph processing.
\newblock In {\em Proc. of SIGMOD '10}, pages 135--146, 2010. ACM.




\bibitem{X10-Brief}
V.~Saraswat, O.Tardieu, D.~Grove, D.~Cunningham, M.~Takeuchi and B.~Herta.
\newblock A Brief Introduction To X10.
\newblock
\url{http://x10.sourceforge.net/documentation/intro/latest/html/}, 2014.

\bibitem{M3R-X10}
A.~Shinnar, D.~Cunningham, V.~Saraswat, and B.Herta. 
\newblock M3R: increased performance for in-memory Hadoop jobs.
\newblock In {\em Proc. of the VLDB Endowment}, 5(12), pages
1736--1747, 2012.

\bibitem{Ullman:2012:DGM:2331042.2331053}
J.~D. Ullman.
\newblock Designing good mapreduce algorithms.
\newblock {\em XRDS}, 19(1):pages 30--34, Sept. 2012. ACM.

\bibitem{Valiant:2011:BMM:1889388.1889509}
L.~G. Valiant.
\newblock A bridging model for multi-core computing.
\newblock {\em Journal of Computer and System Sciences}: 77(1), pages
154--166, Jan. 2011.

\bibitem{Zaharia:2010:SCC:1863103.1863113}
M.~Zaharia, M.~Chowdhury, M.~J. Franklin, S.~Shenker, and I.~Stoica.
\newblock Spark: Cluster computing with working sets.
\newblock In {\em Proc. of HotCloud'10}, pages 10--10, 2010. USENIX Association.


\bibitem{X10PerformanceTuning}
X10 Performance Tuning
\url{http://x10-lang.org/documentation/practical-x10-programming/performance-tuning.html}










\end{thebibliography}


\end{document}